\documentclass[twocolumn,showpacs,preprintnumbers,amsmath,amssymb]{revtex4}

\usepackage{graphicx}
\usepackage{dcolumn}
\usepackage{bm}

\begin{document}

\preprint{APS/123-QED}

\title{Evolution of 4f-Electron States in the Metal-Insulator Transition of PrRu$_4$P$_{12}$}

\author{Kazuaki Iwasa}
\email{iwasa@iiyo.phys.tohoku.ac.jp}
\author{Lijie Hao}
\affiliation{Department of Physics, Tohoku University, Sendai 980-8578, Japan.}
\author{Keitaro Kuwahara}
\author{Masahumi Kohgi}
\author{Shanta Ranjan Saha}%
\altaffiliation[Present address at ]{Muon Science Laboratory, Institute of Material Structure Science, High Energy Accelerator Research Organization,1-1 Oho, Tsukuba, Ibaraki 305-0801, Japan.}
\author{Hitoshi Sugawara}
\altaffiliation[Present address at ]{Department of Mathematical and Natural Sciences, Faculty of Integrated Arts and Sciences, The University of Tokushima, Tokushima 770-8502, Japan.}
\author{Yuji Aoki}
\author{Hideyuki Sato}%
\affiliation{Department of Physics,Tokyo Metropolitan University, Hachioji, Tokyo 192-0397, Japan.}%

\author{Takashi  Tayama}
\author{Toshiro Sakakibara}
\affiliation{Institute for Solid State Physics, University of Tokyo, Kashiwa, Chiba 277-8581, Japan.}%

\date{\today}

\begin{abstract}
Magnetic excitations of the filled skutterudite PrRu$_4$P$_{12}$ exhibiting a metal-insulator (M-I) transition at $T_{\rm M-I}=$ 63 K were studied by inelastic neutron scattering experiment. The spectra at temperatures much lower than $T_{\rm M-I}$ are described as well-defined crystal-field excitations. With approaching $T_{\rm M-I}$, the excitation peaks broaden and shift considerably together with the temperature variation of the carrier number and the atomic displacement in the transition. The 4f-electron state evolve from the well localized state in the insulator phase to the strongly hybridized itinerant state by p-f mixing near $T_{\rm M-I}$. The hybridization is responsible for the M-I transition of PrRu$_4$P$_{12}$.
\end{abstract}

\pacs{71.70.Ch, 71.30.+h, 78.70.Nx}
\maketitle

\section{\label{sec:level1}Introduction\protect\\}

Many rare-earth and actinide compounds have been studied extensively because of their attractive properties such as heavy electrons, unconventional superconductivity, quantum critical behaviors, valence fluctuation etc.
It is commonly accepted that interplay between f and conduction electrons is responsible for these properties. The hybridization effect often gives rise to a quasielastic or broad magnetic excitation spectra instead of well-defined crystal-field (CF) excitations~\cite{HFspectra}. 
The CF spectral width proportional to temperature in rare-earth systems has been interpreted by so-called Korringa-law in which exchange interaction between f and conduction electrons is taken into account~\cite{Korringa_Becker}. Those of dense Kondo-effect or valence fluctuation systems often show $T^{1/2}$-behavior. It was theoretically studied based on the degenerate Anderson model~\cite{theory_Kuramoto,theory_Cox}. To understand the complex magnetic structures of Ce-monopnictides, the effect of p-f mixing between 4f electrons and  p holes of the pnictogens on the CF levels was discussed~\cite{Takahashi_Kasuya}. It succeeded in explaining the anomalously small CF split between a ground state and a excited state. 

Recently, filled skutterudite RT$_4$X$_{12}$ (R = lanthanide and actinide elements, T = transition metal, X = pnictogen) has been found to show various physical properties  involving f electrons~\cite{RareEarth}. In their crystal structure with the space group Im$\bar{\rm 3}$ ($T_h^5$), X atoms form corner shared octahedra. T and R ions are located inside and between the octahedra, respectively. Since R ions are surrounded by 12 X atoms forming an icosahedron, it is suggested that the various properties originate from the p-f mixing~\cite{Harima03}. Among them, PrRu$_4$P$_{12}$ exhibits a metal-insulator (M-I) transition at $T_{\rm M-I}=$ 63 K~\cite{Sekine97}. Electron and X-ray diffraction studies elucidated a superlattice characterized by the wave vector ${\bf q_0} = (1, 0, 0)$ with Pm$\bar{\rm 3}$ ($T_h^1$) below $T_{\rm M-I}$~\cite{Lee01, Lee04, Hao04}. The band structure study proposed that the Fermi surfaces perpendicular to the [1, 0, 0] axes cross the half way between $\Gamma$ and X points to form a cubic-like shape~\cite{Harima02}. Thus, this phase transition is suggested to be a charge-density-wave (CDW) formation owing to a three-dimensional Fermi-surface nesting by ${\bf q_0}$. No anomaly of magnetic susceptibility was observed at $T_{\rm M-I}$~\cite{Sekine97} and the CF ground state was suggested to be non magnetic from the specific heat measurement~\cite{Sekine_C}. However, an isotropic magnetic moment of about 1$\mu_{\rm B}/{\rm Pr}$ is induced under a magnetic field less than 1 T at lower temperature, as presented in this paper. There has been no trace of magnetic ordering. The electrical resistivity turns up below $T_{\rm M-I}$, and shows shoulder-like anomaly around 40 K. From these facts, it is natural to expect a contribution of 4f electrons to the M-I transition. Since the magnetic excitation spectrum by 4f-electron CF levels reflect the carrier state as mentioned above, the 4f-electron role in the M-I transition of PrRu$_4$P$_{12}$ can be investigated by the magnetic excitation measurement.

To understand the 4f-electron state and the anomalous magnetic state in the M-I transition of PrRu$_4$P$_{12}$, we carried out an inelastic neutron scattering measurement. Clear sharp CF excitations of Pr$^{3+}$ 4f$^2$ electrons are observed at temperatures much lower than $T_{\rm M-I}$. It is remarkable that, with increasing carrier number by elevating temperature, the CF excitation peaks shift and broaden drastically. We will show that PrRu$_4$P$_{12}$ undergoes the M-I transition mediated by strongly hybridized state between 4f electrons and carriers. 

\section{\label{sec:level2}Experimental details\protect\\}

A polycrystalline sample was prepared by the Sn-flux method, whose quality is similar to that used in the transport study~\cite{Saha_R_03}. The inelastic neutron scattering experiments were performed at the time-of-flight spectrometers LAM-D (thermal neutron) and LAM-40 (cold neutron) installed in the pulsed neutron facility KENS in KEK, Japan. Scattered neutrons with the energy of 4.59 meV selected by pseudo-mirror-type pyrolithic graphite crystal analyzers were counted by four or seven detectors at LAM-D and LAM-40, respectively. Sample temperatures were controlled between 5 and 70 K by a cryostat with continuous flow of liquid He.

\section{\label{sec:level3}Results and discussion\protect\\}

\begin{figure}[h]
\includegraphics[width=7.5cm]{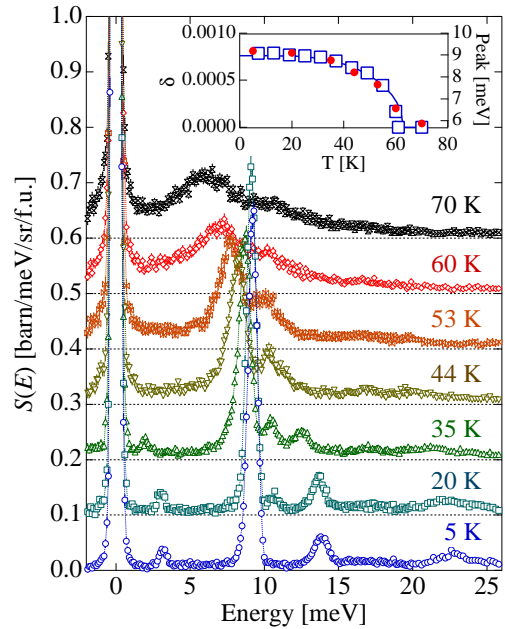}
\caption{\label{f1} Symbols represent $S(E)$ of PrRu$_4$P$_{12}$ measured at LAM-D. Origins of vertical axes of each temperature data are shifted by 0.1. In an inset, open squares depicts temperature dependence of the Ru-ion displacement $\delta$ in units of atomic coordinate determined in the X-ray diffraction study~\cite{Hao04}, a solid line is the BCS gap function fitted to $\delta$, and filled circles are position of the highest inelastic neutron scattering peak.}
\end{figure}

Figure~\ref{f1} represents the temperature dependence of the response functions $S(E)=\frac{1}{N}\left(\frac{k_f}{k_i}\right)^{-1}\frac{d^2\sigma}{d{\Omega}dE_f}$ obtained at LAM-D. These are evaluated by correcting absorption, subtracting the background estimated from the measurement without the sample, and transforming to absolute cross sections by incoherent scattering intensity of vanadium.
At 5 K, clear peaks are seen at 3.2, 9.3, 13.9 and 22.7 meV. Although the width of the last peak is broader than the instrumental resolution in contrast to the other peaks having almost resolution-limited widths, they are presumably attributed to CF excitations, since the scattering-angle dependence of intensity is consistent with that of magnetic form factor of Pr$^{3+}$. It is noticeable that, with increasing temperature, the peaks shift to lower-energy side and become very broad. Such drastic behavior will be discussed based on the strong electron hybridization. Open squares in the inset of Fig.~\ref{f1} depict temperature dependence of Ru-ion displacement $\delta$ determined by the X-ray diffraction study, where the Ru ions are located at $(1/4+\delta, 1/4+\delta, 1/4+\delta)$ and the equivalent positions~\cite{Lee01, Lee04, Hao04}. The maximum-intensity position of inelastic spectra indicated by circles as well as $\delta$ corresponding to the order parameter obey the BCS-type gap function for CDW state shown by a solid line. Then it is natural to consider the CF levels variation coupling strongly with the carrier state and the structural change.

\begin{figure}[h]
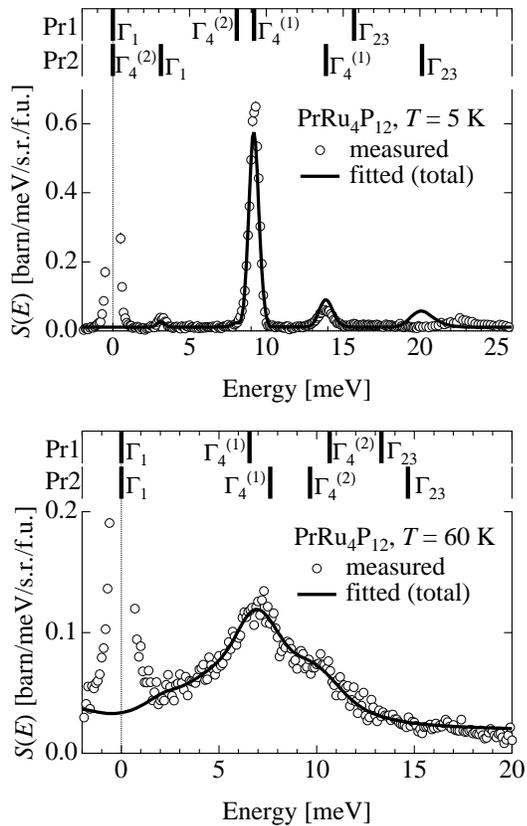

\includegraphics[width=7.5cm]{G_PrRuP_5K_LAMD_2.EPSF}
\includegraphics[width=7.5cm]{G_PrRuP_60K_LAMD_2.EPSF}
\caption{\label{f2} Circles are measured spectra at 5 K (upper part) and at 60 K (lower part). Lines are the fitted results by the model of two CF level schemes. Vertical bars represent CF energies at Pr1 and Pr2 measured from each ground state.
}
\end{figure}

\begin{figure}[h]
\includegraphics[width=7.5cm]{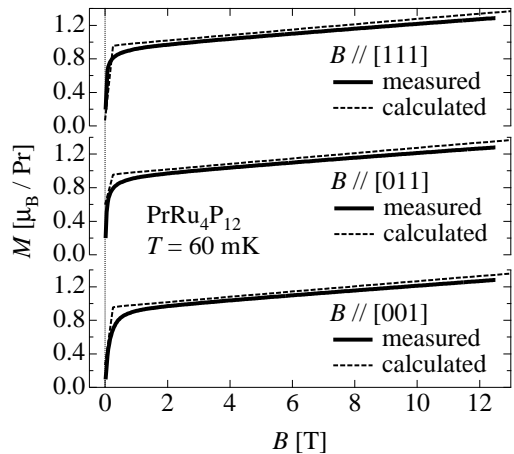}
\caption{\label{f3} Magnetizations measured at 60 mK (solid lines) and those calculated from the CF parameters at 5 K (dotted lines).}
\end{figure}

Because of the superlattice structure below $T_{\rm M-I}$, local environments around the Pr ions at the unit-cell origin and at the body center are different. Therefore, there should be two level schemes for each Pr-ion sites (Pr1 and  Pr2) below $T_{\rm M-I}$. The CF Hamiltonian for 4f electrons in both of Im$\bar{\rm 3}$ and Pm$\bar{\rm 3}$ structures is represented as 
\begin{eqnarray}
{\cal H}_{\rm CF} = A_4(O_4^0 + 5O_4^4) &+& A_{6}^{c}(O_6^0 - 21O_6^4) \nonumber \\
&+& A_{6}^{t}(O_6^2 - O_6^6), 
\label{eq1}
\end{eqnarray}
where $O_m^n$ expresses a Stevens' operator equivalent~\cite{Takegahara}. The last term of $A_{6}^{t}$ is caused by the lack of the point symmetry $C_4$ in $T_h$. The 4f$^2$-electron state of Pr$^{3+}$ split to four levels; a singlet $\Gamma_1$, a non-magnetic non-Kramers doublet $\Gamma_{23}$, and triplets $\Gamma_4^{(1)}$ and $\Gamma_4^{(2)}$.  We carried out a least-squares fitting procedure of calculated CF excitation spectra in the full measured energy range to the data of $S(E)$ at 5 K. We took free parameters of the CF coefficients in eq.~\ref{eq1}, a scale factor for a ratio of calculated cross section to the experimental result, and a constant background to approximate phonon contribution which is very small compared with the CF excitation peaks. The peak widths are assumed to be equal to the instrumental resolution. The result shown by solid lines in the upper figure of Fig.~\ref{f2} reproduces well the experimental data shown by circles, except the broad peak observed at 22.7 meV. 
The best fit CF coefficients are shown in Table~\ref{t1}.
\begin{table*}
\caption{\label{t1} CF coefficients and scale factors obtained from the analysis of $S(E)$.
}
\begin{ruledtabular}
\begin{tabular}{cccccccc}
Parameter & 5 K & 20 K & 35 K & 44 K & 53 K & 60 K & 70 K  \\ \hline
$A_4$(Pr1) [mK] & -6.816 & -5.571 & -16.69  & -14.10 & -12.90 & -17.17  & -15.06\\
$A_{6}^{c}$(Pr1) [mK] & 0.9505 & 0.952 & 0.860  & 0.879 & 0.844 & 0.715   & 0.729\\
$A_{6}^{t}$(Pr1) [mK] & 0.46 & 0.41 & 2.01  & 2.56 & 2.36 & 5.69   & 7.31\\
$A_4$(Pr2) [mK] & 41.2 & 42.7 & 22.25  & 17.45 & 13.83 & -11.50  &  -15.06\\
$A_{6}^{c}$(Pr2) [mK] & 1.41 & 1.45 & 0.944  & 0.916 & 0.794 & 0.847  & 0.729\\
$A_{6}^{t}$(Pr2) [mK] & 10.0 & 8.07 & 7.50  & 7.43 & 8.68 & 3.09  & 7.31\\
scale & 0.940 & 1.001 & 0.850 & 0.888 & 0.986 & 1.067 & 1.227\\
\end{tabular}
\end{ruledtabular}
\end{table*}
It is characteristic that, in lower temperature region, the ground state of Pr1 is non-magnetic $\Gamma_1$ and the first excited state magnetic $\Gamma_4^{(2)}$, and vice versa for Pr2. In order to check the validity of the obtained CF scheme, we calculated magnetization based on the CF coefficients at 5 K. As shown in Fig.~\ref{f3}, the resultant isotropic magnetization dominated by the $\Gamma_4^{(2)}$ ground state of Pr2 agrees quite well with the experimental data at 60 mK. 
The discrepancy between the observation and the calculation at low fields can be dissolved by taking into account a split of the $\Gamma_4^{(2)}$ expected from the observed Schottky peak of specific heat at 0.3 K, although the origin of the split is unknown.
Because the nearest-neighbor Pr ions (Pr1) of Pr2 are non-magnetic, no magnetic ordering is naturally expected. The specific heat value of 0.51 J/K/mol at 9 K evaluated from the CF parameters at 5 K, corresponding to an excitation from $\Gamma_4^{(2)}$ to $\Gamma_1$ of Pr2, is close to the observed value of 0.6 J/mol/K~\cite{Sekine_C}. Such quantitative agreement with the magnetization and the specific heat supports the present assignment of the CF levels.

\begin{figure}[h]
\includegraphics[width=7.5cm]{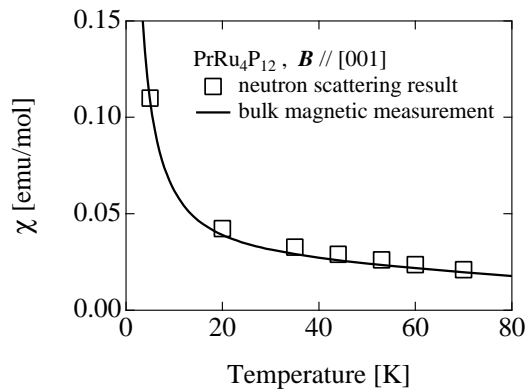}
\caption{\label{f4} A solid line represents measured bulk magnetic susceptibility under magnetic field applied along the [001] axis, and open squares the calculated one based on the CF parameters determined in the present neutron scattering experiments.}
\end{figure}

For analyzing the data at higher temperatures, the CF peaks are assumed to be expressed by Lorentzians with a finite width adjusted in the fitting procedure involving convolution of the instrumental resolution. Since the two Pr-ion sites are equivalent above $T_{\rm M-I}$, we analyzed the data at 70 K under constraint of the same CF parameters for Pr1 and Pr2. As shown in the lower part of Fig.~\ref{f2}, the fitted curve agrees well with the experimental result at 60 K. The resultant parameters are listed in Table~\ref{t1}.
The values of CF Hamiltonian coefficients do not show a smooth dependence on temperature. This is due to following reasons. The cross section of the excitation from the ground state $\Gamma_1$ to $\Gamma_4^{(2)}$ at 8.14 meV of Pr1 at 5 K is about 3\% of that from $\Gamma_1$ to $\Gamma_4{(1)}$ at 9.27 meV, which were resolved only in the higher-resolution measurement at LAM-40. The cross section of the excitation from $\Gamma_4^{(2)}$ to $\Gamma_1$ at Pr2 is also small. These smaller peaks are hidden at high temperature region because of the spectral broadening. Thus, the determination of the small-peak position and intensity shown in Fig.~\ref{f1} contains uncertainty. However, the overall spectra are well fitted and the scale factors are almost one as tabulated in Table~\ref{t1}. We compared also the temperature variation of bulk magnetic susceptibility with that of the CF schemes determined in the present neutron scattering study. As seen in Fig.~\ref{f4}, the magnetic susceptibility measured under magnetic field of 0.1 T applied along the [001] axis (a solid line) agrees quite well with that calculated from the presently determined CF Hamiltonian parameters as well as the Zeeman term (open squares). 
Thus, the present analysis succeeds in reproducing the experimentally determined cross sections quantitatively~\cite{Thalmeier}.

\begin{figure}[h]
\includegraphics[width=7.0cm]{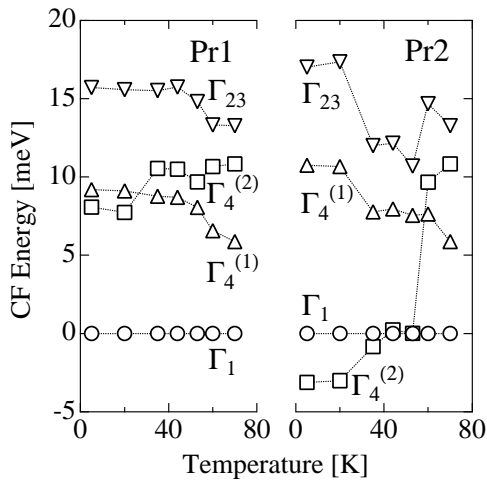}
\caption{\label{f5} CF level schemes of Pr1 and Pr2 as functions of temperature. The energy is defined as difference from the $\Gamma_1$ level.}
\end{figure}

\begin{figure}[h]
\includegraphics[width=6.5cm]{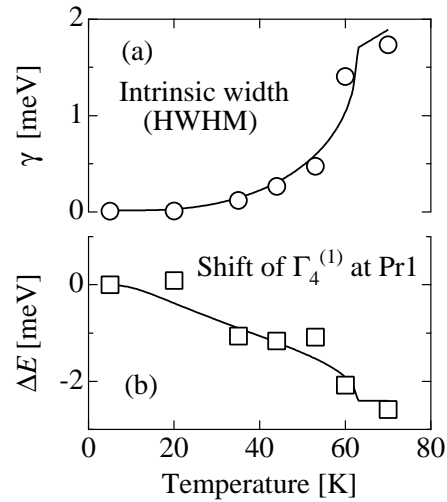}
\caption{\label{f6} (a) Circles are the HWHM of CF peaks and a line calculated one based on eqs.~2 and 3. (b) Squares are the shift of eigenenergy of $\Gamma_4^{(1)}$ at Pr1 and a line based on eq.~\ref{eq3}.}
\end{figure}

Figure~\ref{f5} shows the resultant temperature variation of the level schemes of Pr1 and Pr2, in which the zero energy is at the $\Gamma_1$ level.
The non-magnetic ground state $\Gamma_1$ of both Pr1 and Pr2 around $T_{\rm M-I}$ is consistent with the suggestion from magnetic susceptibility and specific heat data~\cite{Sekine97, Sekine_C}. The ground state of Pr2 switches to $\Gamma_4^{(2)}$ below about 40 K. This switch may give rise to the plateau of electrical resistivity around this temperature~\cite{Sekine97}. As depicted by symbols in Fig.~\ref{f6}~(a), the intrinsic half width at half maximum (HWHM) varies from $\sim0$ at 5 K to 1.73 meV at 70 K. This unusual CF evolution evidences a strong interaction of 4f electrons with carriers. 

The observed peak broadening will be analyzed by the following form of the HWHM $\gamma(T)$ up to $T \cong 6\Delta_{\rm CF}$, where $\Delta_{\rm CF}$ is the CF split energy, expressed as
\begin{eqnarray}
\gamma(T) = \gamma(0) + 4\pi\{J_{\rm ex}(g_J-1)N(E_{\rm F})\}^2k_{\rm B}T, 
\label{eq2}
\end{eqnarray}
where $J_{\rm ex}$ denotes an exchange integral between 4f and conduction electrons and $N(E_{\rm F})$ the density of states at the Fermi level~\cite{Korringa_Becker}. In contrast to conventional metals  showing $T$-linear behavior of $\gamma(T)$ due to temperature-independent $N(E_{\rm F})$, the  broadening of CF peaks in PrRu$_4$P$_{12}$ is strongly non-linear against $T$ as depicted in Fig.~\ref{f6}~(a). This phenomenon can be ascribed to the variation of electronic state of PrRu$_4$P$_{12}$ in the M-I transition. We analyze the data based on eq.~\ref{eq2} by simply assuming that the $N(E_{\rm F})$ is proportional to the carrier number $n(T)$. It is presumed to be expressed as
\begin{eqnarray}
n(T) = n_0e^{-\Delta_g(T)/k_{\rm B}T}, 
\label{eq3}
\end{eqnarray}
where $\Delta_g(T)$ is an activation energy following the BCS gap function reproducing the atomic displacement below $T_{\rm M-I}$ as shown in the inset of Fig.~\ref{f1}. $\Delta_g(0) =$ 37 K is evaluated from the electrical resistivity~\cite{Sekine97}. Calculated $\gamma(T)$ by substituting $n(T)$ for $N(E_{\rm F})$ and by putting arbitrary values of $\gamma(0) = 0.016
~{\rm meV}$ and $J_{\rm ex}n_0 = 0.23$ reproduces the experimental HWHM well, as shown by a line in Fig.~\ref{f6}~(a). The success of this simple analysis supports that the hybridization effect becomes significant with increasing temperature near $T_{\rm M-I}$.

The shift of CF excitation peaks is also explained by the carrier-state variation in the M-I transition. The p-f mixing forces the valence band level to be raised and the mixed CF level to be lowered~\cite{Takahashi_Kasuya}. The CF-energy shift is proportional to the number of holes in valence band. In the case of PrRu$_4$P$_{12}$, the holes in the 49th band is thought to be responsible for the p-f mixing~\cite{Harima02}. As it is reduced below $T_{\rm M-I}$, the mixed CF level energy is expected to increase. In addition, to keep the center of gravity of CF energy, even the CF levels with less mixing also shift. In Fig.~\ref{f6}~(b), the experimentally determined shift of eigenenergy of $\Gamma_4^{(1)}$ at Pr1 is shown as a typical example, together with a line of $-n(T)$ in eq.~\ref{eq3} with an arbitrary value of $n_0=2.4$. Their temperature variations are similar to each other, so that the strong hybridization between f-electron orbit and p-electron holes at high temperature region is again supported.

\section{\label{sec:level3}Summary\protect\\}

It is concluded that the M-I transition of PrRu$_4$P$_{12}$ is a new type of CDW transition mediated by the strong hybridization between 4f-electron and carrier states. Above $T_{\rm M-I}$, the instability of the p-hole band due to the three dimensionally perfect nesting is enhanced by the high density of state at the Fermi  level owing to the hybridization effect, as evidenced by the present work. It gives rise to the CDW transition resulting in lowering of the band energy by the gap formation and loosing the hybridization with 4f-electron state. This scenario can explain the observed evolution of 4f-electron state through the transition, and explain also the reason why LaRu$_4$P$_{12}$ without 4f electron does not undergo a CDW transition in spite of similar nesting conditions as that of  PrRu$_4$P$_{12}$~\cite{Harima_LaRuP}.

\begin{acknowledgments}
The authors acknowledge greatly Prof.~H.~Harima and Dr.~K.~Matsuhira for their fruitful discussions and the neutron-scattering instrument group LAM at KEK for the experimental assistance. This study is partially supported by the Grants-in-Aid for Scientific Research from Ministry of Education, Culture, Sports, Science and Technology, Japan (Young Scientists (B) (No.~15740219) and Scientific Research Priority Area ``Skutterudite" (No.~15072206)).
\end{acknowledgments}


\begin{thebibliography}{9}

\bibitem{HFspectra} E. Holland-Moritz and G. H. Lander, 
\textit{Handbook on the Physics and Chemistry of the Rare Earths Vol. 19}  (Elsevier Science B. V., Amsterdam 1994) p.1.
\bibitem{Korringa_Becker} K. W. Becker, P. Fulde and J. Keller, 
Z. Physik B ${\bf 28}$, (1977) 9.
\bibitem{theory_Kuramoto} Y. Kuramoto and E. M\"uller-Hartmann,
J. Magn. Magn. Mater. ${\bf 52}$ 122 (1985).
\bibitem{theory_Cox} D. L. Cox, N. E. Bickers and J. W. Wilkins, 
J. Magn. Magn. Mater. ${\bf 54-57}$, 333 (1986).
\bibitem{Takahashi_Kasuya} H. Takahashi and T. Kasuya,
J. Phys. C, Solid State Phys. {\bf 18}, 2697 (1985); {\bf 18}, 2709 (1985); {\bf 18}, 2721 (1985); {\bf 18}, 2731 (1985); {\bf 18}, 2745 (1985); {\bf 18}, 2755 (1985).
\bibitem{RareEarth} B. C. Sales, \textit{Handbook on the Physics and Chemistry of Rare Earths Vol . 33} (Elsevier Science B. V., Amsterdam, 2003) p.1.
\bibitem{Harima03} H. Harima and K. Takegahara,
J. Phys.: Condens. Matter ${\bf 15}$, S2081 (2003).
\bibitem{Sekine97} C. Sekine, T. Uchiumi, I. Shirotani and T. Yagi, 
Phys. Rev. Lett. ${\bf 79}$, 3218 (1997).
\bibitem{Lee01} C. H. Lee, H. Matsuhata, A. Yamamoto, T. Ohta, H. Takazawa, K. Ueno, C. Sekine, I. Shirotani and T. Hirayama, 
J. Phys.: Condens. Matter ${\bf 13}$, L45 (2001).
\bibitem{Lee04} C. H. Lee, H. Matsuhata, H. Yamaguchi, C. Sekine and I. Shirotani, 
J. Magn. Magn. Mater. ${\bf 272-276}$, 426 (2004).
\bibitem{Hao04} L. Hao, K. Iwasa, K. Kuwahara, M. Kohgi, S. R. Saha, H. Sugawara, Y. Aoki, H. Sato, C. Sekine, C. H. Lee and H. Harima, 
J. Magn. Magn. Mater. ${\bf 272-276}$, e271 (2004).
\bibitem{Harima02} H. Harima and K. Takegahara,
Physica B ${\bf 312-313}$, 843 (2002).
\bibitem{Sekine_C} C. Sekine, T. Inaba, I. Shirotani, M. Yokoyama, H. Amitsuka, T. Sakakibara, 
Physica B ${\bf 281\&282}$, 303 (2000).
\bibitem{Saha_R_03} S. R. Saha, H. Sugawara, T. Namiki, Y. Aoki and H. Sato, 
J. Phys.: Condens. Matter ${\bf 15}$, S2163 (2003).
\bibitem{Takegahara} K. Takegahara, H. Harima and A. Yanase, 
J. Phys. Soc. Jpn. ${\bf 70}$, 1190 (2001); ${\bf 70}$, 3468 (2001); ${\bf 71}$, 372 (2002).
\bibitem{Thalmeier}
The analytical form of CF eigenenergies for $T_h$ symmetry~\cite{Takegahara} does not reproduce the peak at 22.7 meV observed below 20 K. The discrepancy may be explained by a coupling between the CF states and optical phonons distorting the CF potential, as the bound state in CeAl$_2$ studied by P. Thalmeier and P. Fulde, Phys. Rev. Lett. ${\bf 49}$, 1588 (1982). We expect a high density of states of optical phonons near the $\Gamma_{23}$ level at Pr2 of PrRu$_4$P$_{12}$, which couple to the excitation from $\Gamma_4^{(2)}$ to $\Gamma_{23}$. However, the coupling has no direct influence on the thermal evolution of the CF schemes and should be ignored here. 
\bibitem{Harima_LaRuP} H. Harima,
Prog. Theor. Phys. Suppl. ${\bf 138}$, 117 (2000).


\end{thebibliography}

\end{document}